%% file: 0_main.tex
\documentclass[fleqn,10pt]{wlscirep}
\usepackage{color}

\usepackage[normalem]{ulem}

\usepackage{bm}
\usepackage{inputenc}
\usepackage{amsmath}
\usepackage{subcaption}
\usepackage{cleveref}

\usepackage{comment}
\usepackage{appendix}

\title{Improving the visibility of minorities through network growth interventions}

\author[1,*]{Leonie Neuhäuser}
\author[2,*]{Fariba Karimi}
\author[2,3]{Jan Bachmann}
\author[4,5,2]{Markus Strohmaier}
\author[1,*]{Michael T. Schaub}

\affil[1]{Computational Network Science Group,
RWTH Aachen University,
Germany}
\affil[2]{Complexity Science Hub Vienna, Austria}
\affil[3]{Department of Network and Data Science, Central European University, Vienna, Austria}
\affil[4]{University of Mannheim, Germany}
\affil[5]{GESIS - Leibniz Institute for the Social Sciences, Germany}
\affil[*]{neuhaeuser@cs.rwth-aachen.de, karimi@csh.ac.at, schaub@cs.rwth-aachen.de}

\begin{abstract}
Improving the position of minorities in networks via interventions is a challenge of high theoretical and societal importance.
In this work, we examine how different network growth interventions impact the position of minority nodes in degree rankings over time. 
We distinguish between two kinds of interventions: 
(i) \emph{group size} interventions, such as introducing quotas, that regulate the ratio of incoming minority and majority nodes; and
(ii) \emph{behavioural} interventions, such as homophily, i.e. varying how groups interact and connect to each other. 
We find that even extreme group size interventions do not have a strong effect on the position of minorities in rankings if certain behavioural changes do not manifest at the same time. 
For example, minority representation in rankings is not increased by high quotas if the actors in the network do not adopt homophilic behaviour. 
As a result, a key finding of our research is that in order for the visibility of minorities to improve, group size and behavioural interventions need to be coordinated. 
Moreover, their potential benefit is highly dependent on pre-intervention conditions in social networks. 
In a real-world case study, we explore the effectiveness of interventions to reach gender parity in academia. 
Our work lays a theoretical and computational foundation for further studies aiming to explore the effectiveness of interventions in growing networks.

\end{abstract}

\begin{document}
\flushbottom
\maketitle

\thispagestyle{empty}

\input{1_Introduction.tex}

\input{2_Results.tex}
\input{3_Discussion.tex}
\input{4_Methods.tex}

\urlstyle{same}
\bibliography{bib}
\end{document}

%% file: 1_Introduction.tex
\section{Introduction}
\label{sec:Introduction}

Historical disadvantages can result in the marginalisation of social groups.
While there are many different mechanisms at play, research has shown that the social network structure plays a vital role in the creation of structural inequalities and marginalisation\cite{dimaggio2012network,karimi2022minorities}.
For example, the position of group members in a network determines their access to information \cite{dimaggio2012network}, social capital \cite{van2015inequality}, and their position in algorithmic rankings \cite{karimi_homophily_2018,espin-noboa_inequality_2022}.
However, societies and their social networks change and evolve over time: as a result of institutional interventions, societal opportunities, or behavioural change, more people from disadvantaged backgrounds may enter certain social settings.
For example, in STEM fields People of Color or women have been historically underrepresented \cite{national_science_foundation_national_center_for_science_and_engineering_statistics_women_2015, eccles_bringing_1989}. 
To combat those historical disadvantages, various intervention measures, e.g., scholarship programs, affirmative action, and other support structures \cite{eccles_bringing_1989, armstrong_starting_2015} have been put in place.
Moreover, changes in behaviour can impact how new arriving individuals connect to the network as social attitudes and level of mixing between the groups may change over time \cite{goodreau_birds_2009}. 

One form of change that we find in real-world social networks is that the size of the marginalised group increases over time. 
For example, in the academic context, measures such as scholarship programs, specific funds for hiring faculty from marginalised groups or affirmative action policies \cite{eccles_bringing_1989} have already been put in place.
Moreover, migration or political instabilities can lead to an increase of diversity in a country's workforce \cite{forsyth_how_2015}. 
In the business environment, various countries, for example Israel \cite{izraeli_paradox_2000} and Norway \cite{seierstad_for_2011}, have enacted laws
that favour the appointment of women in corporate boards of directors \cite{sojo_reporting_2016}.
Even though not all of these changes are enacted from an external party in a controlled manner, we refer to these types of changes collectively as \textit{group size interventions} in the following.



However, merely increasing the group sizes of minorities without considering the social network structure and the position of social groups is insufficient to resolve structural inequalities. From the network-theoretic prospective, the visibility of minority groups in a network is dependent on both their relative sizes and the strength of group mixing biases \cite{karimi_homophily_2018,oliveira2022group}. Homophily, ``similarity breeds connection'' is one of those fundamental tie formation mechanisms that leads a higher tendency for ingroup mixing and often  associated closely with attributes such as gender, race, and ethnic backgrounds\cite{mcpherson_birds_2001}. This interplay of homophily and group size indicates that size alone does not determine the marginalisation of a group and increasing the group size of a minority (i.e., increase participation of a minority) can thus not be the only intervention to combat inequality. Indeed, recent findings highlight that even if the group size increases, depending on the homophily, women as a minority group can remain under-represented in directed citation networks \cite{nettasinghe_emergence_2021}. Beyond group size interventions, it is thus necessary to investigate how changes in people's behaviour affect their visibility in social networks. We call this form of intervention that is derived from individual's behaviour and operates through homophily as \textit{behavioural intervention}.

Behavioural interventions can be manifested in various ways. For example, supervisors or top-ranked executives may realise that diversity and inclusion can increase innovative outcomes and revenues \cite{herring_does_2009} and decide to interact with and hire more people of diverse backgrounds.
Changing the mixing preferences can also naturally arise when minority sizes change: research has shown that students of low socioeconomic status start to interact in a more homophilic way as soon as their numbers increase~\cite{alvarez-rivadulla_college_2022}.
Moreover, newcomers may be advised or mentored by individuals in their own communities through variety of support networks \cite{dennissen_diversity_2019}. 
While addressing the individual's career development and community building, such support networks are much less articulate about removing the barriers to inclusion in the organisation \cite{dennissen_diversity_2019}. This underlines that behavioural interventions have to be combined with institutional changes to increase their effectiveness. From the network-theoretic prospective there is a mathematical limit in which minorities can enhance their network visibility. When minorities are numerically small, the majority have the power and resources to make effective network changes \cite{oliveira2022group}. In other words, behavioural change without institutional enforcement may not be enough to decrease inequality in visibility.


Altogether, the previous examples show that not all types of interventions are successful, and network interventions in isolation often cannot combat structural inequalities. Indeed, computational models and quantitative methods are needed to go beyond only identifying existing inequalities and evaluate the impact of different types of interventions in evolving social systems \cite{noauthor_broader_2022}.
Despite its importance, a rigorous and quantitative investigation of the interplay between group size and behavioural interventions on overcoming structural barriers and  marginalisation is still missing.


In this work, we examine the effect of interventions in growing networks and their effects on the ranking visibility of minorities.
To this end, we devise a two-phase network model in which homophily and group sizes are time-dependant.
We explore two kinds of interventions: (i) \emph{group size} interventions, which regulate the ratio of incoming minority and majority nodes and (ii) \emph{behavioural} interventions, which vary the homophily of nodes (group mixing) regulating how minority and majority nodes link to each other.
We study the impact of these interventions on the network position of the minorities in top degree ranks.
Our analysis enables us to evaluate the effectiveness of those intervention policies that aim to correct or overcome the structural inequalities in social networks.
Finally, we investigate ``what-if'' scenarios in the academic setting by exploring the effectiveness of certain interventions to reach gender parity in academia under our model.
Our results demonstrate that group size and behavioural interventions have to be coordinated to be effective and are highly dependent on the pre-intervention network conditions they are building on.


While there are various dimensions of marginalisation, we focus on inequality that emerges from and is exacerbated by network structure. 
In social networks, centrality measures such as degree centrality determine the social capital of individuals \cite{li_co-authorship_2013} and the perception of minorities \cite{lerman_majority_2016,lee_homophily_2019}. 
Individuals in top centrality ranks are more visible in ranking and recommender algorithms \cite{espin-noboa_inequality_2022} and they have better access to resources that are available in social networks \cite{sparrowe_social_2001}. The degree ranking of minority groups in a network is a function of relative group sizes and the presence or absence of homophilic behaviour \cite{karimi_homophily_2018}. This follows the logic that homophily restricts the minority group's ability to establish links with a majority group which then leads to a disadvantage for this group in degree-based rankings.  
However, the majority of works on modelling growing social networks assume that people of different groups maintain their group mixing biases or there is no change in their proportion over time. 
Indeed, recent studies have found that time dependencies in interactions can significantly alter network results and dynamics, e.g., impact communities in networks \cite{mucha_community_2010} and alter diffusion of information \cite{holme_modern_2015,scholtes_causality-driven_2014,neuhauser_consensus_2021}.

More importantly, in large scale-free social networks driven by a preferential attachment mechanism, early arriving individuals in top ranks stabilise their position and give little opportunities to others to reach those ranks \cite{ghoshal_ranking_2011}. 
In such cases, accumulated early advantages may create structural barriers for minorities to enhance their social capital.
Therefore, it is essential to take into account group size changes and behavioural changes in the network growth process to investigate how network positions of groups are formed and what kinds of interventions can improve the position of disadvantaged groups in the future. 
Network interventions have previously been considered as a process that utilises social networks to accelerate behaviour change or improve organisational performance \cite{valente_network_2012}.
For example, when diffusion between groups is expected to be difficult, bridging individuals or rewiring the connectivity may be an effective strategy \cite{valente2010bridging,centola2021change}. 
Our work here extends this stream of literature by introducing and systematically analysing network growth interventions.

%% file: 2_Results.tex
\section{Results}
\label{sec:results}
\subsection*{Conceptual two-phase network model}


We explore the change of network visibility of a minority group as an effect of growth interventions in a simple yet systematic way.
We construct a two-phase growth model as an extension of the \textit{BA-Homophily}~\cite{karimi_homophily_2018} model which itself is an extension of the well known Barab\'{a}si-Albert (\textit{BA}) preferential attachment model \cite{barabasi_emergence_1999}.
In the original \textit{BA} model \cite{barabasi_emergence_1999}, new nodes arrive iteratively to the system and connect to existing nodes with a probability proportional to their degree (preferential attachment).
This creates a ``rich get richer'' phenomenon because large degree nodes receive more new links.
The \textit{BA-Homophily}~\cite{karimi_homophily_2018} model extends the BA model, such that each node is now assigned to one of two differently sized groups.
A node can either belong to the minority group, or the majority group. 
For every new node entering the system, the attachment probability now depends not only on the node's degree, but also on the group labels of the respective nodes.
The impact of the group structure is determined by the minority group size $min$ and a homophily parameter $h$.
The homophily parameter varies between $0$ and $1$ and regulates the probability that nodes that belong to the same group attach.
For example, $h=0.1$ describes a very heterophilic setting where nodes of different groups have a higher probability of connecting, whereas $h=0.9$ describes a system with preferred attachment to the same group, i.e., homophily.
The case of $h=0.5$ describes a lack of group preference and recovers the original Barab\'{a}si-Albert model.
Details of the model are described in Section \ref{sec:methods}.

\begin{figure*}
    \centering
    \includegraphics[width=0.9\textwidth]{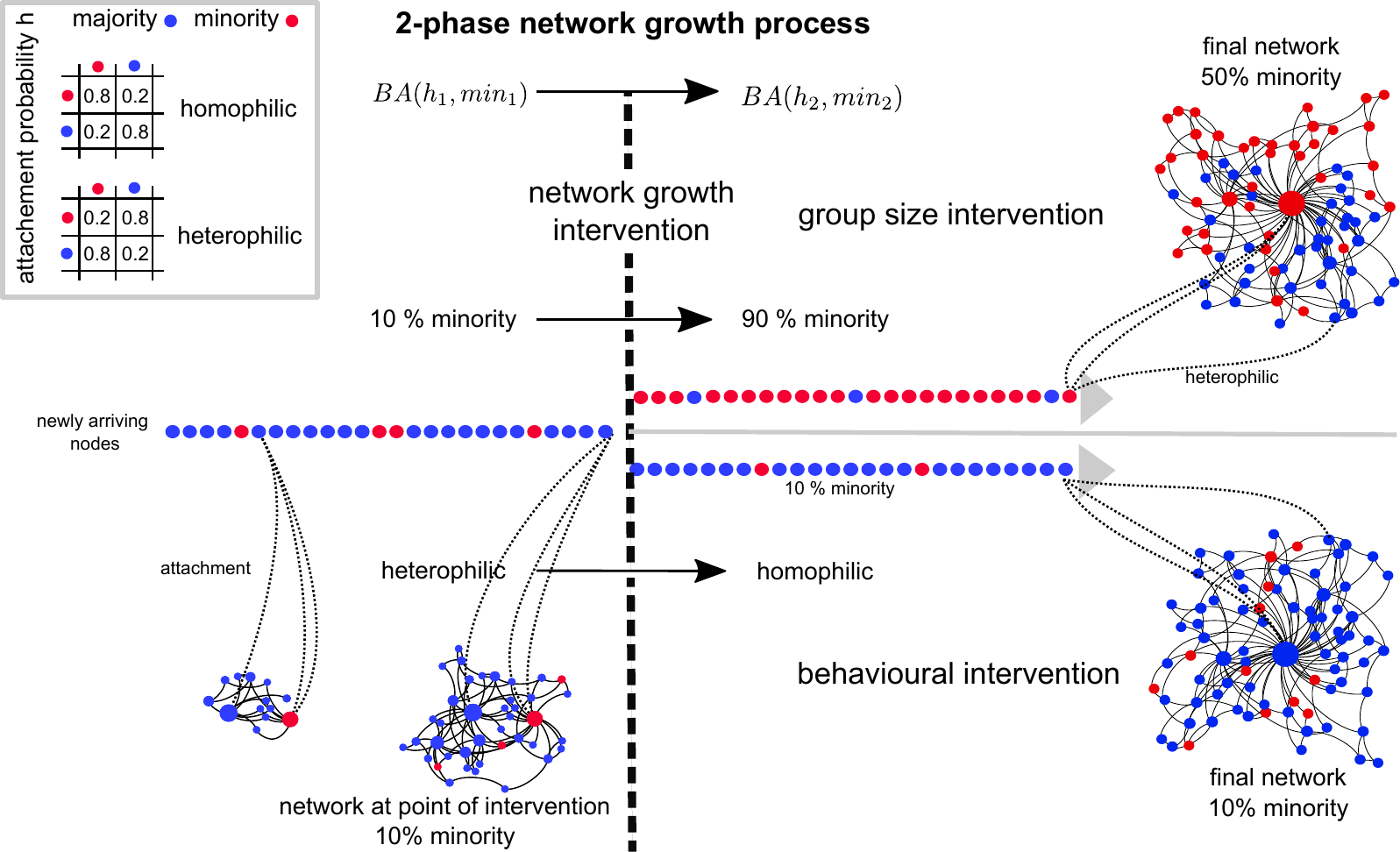}
   \caption{\textbf{Modelling interventions with a two-phase network growth model.} 
   We consider attributed networks with two groups, a majority and a minority. 
   The network growth process, $BA_h(h,min)$, is driven by preferential attachment, and a tunable homophily ($h$) and minority group size ($min$). 
   Changing the two parameters at a certain intervention time point results in a two-phase process. 
   The network growth in the pre-intervention phase happens according to $BA_h(h_1,min_1)$. 
   The network growth in the second phase is determined by $BA_h(h_2,min_2)$, where we assume $min_2 \geq min_1$. 
   For two growth phases with equal length, this results in a final network with a total minority size of  $min_{total} = (min_2+min_1)/2$. 
   A network growth intervention is thus fully defined by $BA_h(h_1,min_1) \rightarrow BA_h(h_2,min_2)$. 
   We can distinguish two different types of interventions. 
   (i) We interpret the change in minority fraction in the incoming nodes as group size intervention, materialising in a quota for the newly arriving nodes. (ii) The switch in the homophily parameter, which determines if nodes like to attach to their kind, is interpreted as a behavioural intervention. 
   }
    \label{fig:1}
\end{figure*}

A novel contribution of this work is to propose a two-phase network growth model that mimics two important social processes. 
First, the formation of structural inequalities that emerge in social networks due to certain pre-existing societal biases \cite{karimi2022minorities} (homophily and minority size in the pre-intervention phase).
Second, the effect of different interventions on changing those initial structural inequalities.  
We do so by modelling network interventions via time-dependent parameters in the network growth process, as shown in Figure \ref{fig:1}. 

Changing the model parameters at a certain intervention time point results in a two-phase process. 
The network growth in the \textit{pre-intervention} phase happens according to initial minority size ($min_1$) and initial homophily ($h_1$), $BA_h(h_1,min_1)$. 
Throughout this phase, $N_1$ nodes join the network. 
Starting with the network that results from this process, the network growth in the second phase, \textit{post-intervention}, is determined by $BA_h(h_2,min_2)$.
During this phase $N_2$ nodes join, resulting in a final network of $N=N_1+N_2$ nodes.

Without restricting ourselves, in this paper we consider two growth phases with the same length for simplicity, i.e., the same amount of nodes joins in the pre- and post-intervention phase, ($N_1=N_2$).
Further, we focus on growth of the minority in the second phase, i.e. $min_2 \geq min_1$.
The two-phase growth process then results in a final network with a total minority size of  $min_{total} = (min_1+min_2)/2$. 
A network growth intervention is thus fully defined by the change of model parameters $BA_h(h_1,min_1) \rightarrow BA_h(h_2,min_2)$. 

Details related to the properties of the model can be found in Section \ref{sec:methods}. 
Note that in a real-world context, both group size and behavioural interventions can be of exogenous or endogenous nature. 
They may happen as a result of institutions establishing policies to protect and encourage groups, due to external events, or as a result of a natural change of the population and preference changes of individuals. 
In this paper we do not distinguish why the interventions happen, but simply examine the resulting outcomes.

\subsection*{Minority representation in rankings}
\begin{figure*}
    \centering
    \begin{subfigure}[b]{\textwidth}
        \centering
        \includegraphics[width=\textwidth]{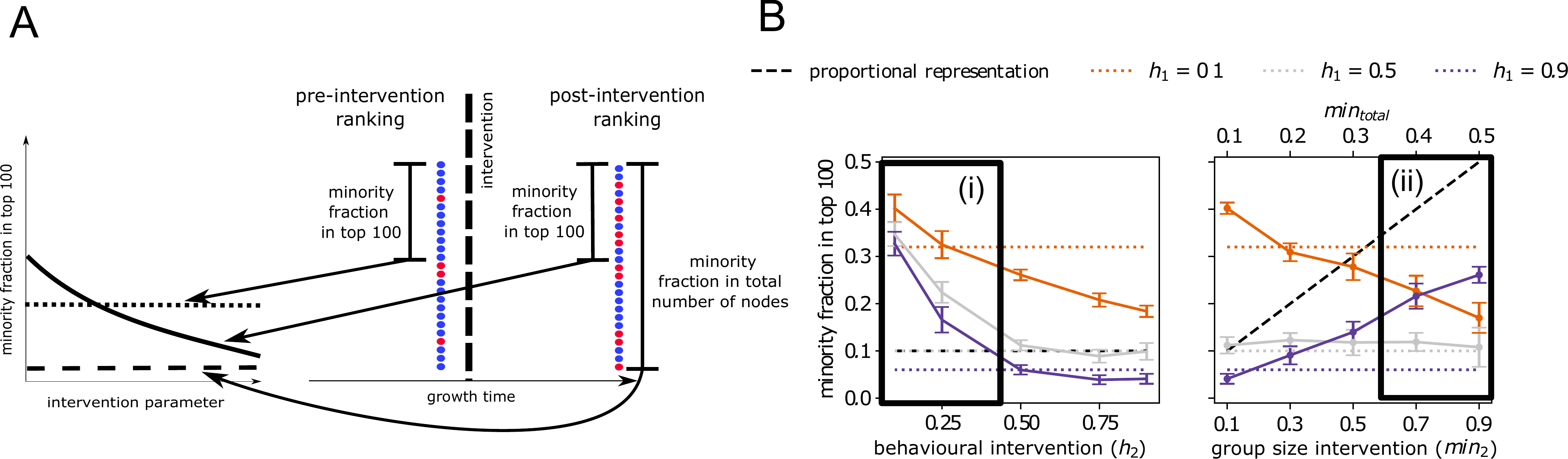}
    \end{subfigure}
    \caption{\textbf{The effects of behavioural and group size interventions on minority representation in rankings.} \\\textbf{A} Schematic description of how we evaluate the effects of network growth interventions on minority representation in rankings. Considering a network growth intervention $BA_h(h_1,min_1) \rightarrow BA_h(h_2,min_2)$, rankings are inferred from the node degrees from the pre-intervention network (at the point of intervention) and from the post-intervention (final) network. 
    We then compare the fraction of the minority within the top 100 nodes of these rankings. 
    We compare the pre-intervention network (dotted line), generated by $BA_h(h_1, min_1)$, with the post-intervention network (solid line), generated by $BA_h(h_2, min_2)$. 
    Here, one of the two intervention parameters, $h_2$ or $min_2$, is varied on the x-axis to evaluate the effect of different values. The dashed black line indicates what would be a proportional representation according to the total minority fraction of the final post-intervention network. 
    \textbf{B} Evaluation of the separate impact of behavioural interventions and group size interventions.  In the left figure, we first evaluate the impact of a behavioural intervention $h_1 \rightarrow h_2$ on the minority representation. We fix the total minority fraction of $min_1=min_2=0.1$  and thus consider three intervention settings: $h_1=0.2$ (heterophilic pre-intervention network, orange), $h_1=0.5$ (homophily-neutral pre-intervention network, grey) and $h_1=0.9$ (homophilic pre-intervention network, purple). 
    We then vary $h_2$ between heterophilic ($h_2=0.1$) and homophilic ($h_2=0.9$) preferences.  For all types of pre-intervention networks, the minority fraction in the top nodes increases for heterophilic post-intervention behaviour (i). This effect decreases as a function of $h_2$. The three settings mainly differ in their baseline of the change:  Starting from a heterophilic pre-intervention network (orange), the minority is initially over-represented in top-ranked nodes. 
    Changing to homophilic post-intervention behaviour lowers the minority's visibility in comparison to the pre-intervention baseline, but the general over-representation in comparison with the proportional representation remains. In contrast, for a homophilic and random initial setting,  the minority is under-represented in the pre-intervention rankings. Changing from homophilic to heterophilic behaviour greatly benefits the minority (i), reversing them from under- to over-representation. \textbf{B} right: we evaluate the impact of group size interventions by fixing the mixing preferences $h_1 = h_2$ and thus examining $BA_h(0.1,0.1) \rightarrow BA_h(0.1,min_2)$ (heterophilic setting), $BA_h(0.5,0.1) \rightarrow BA_h(0.5,min_2)$ (neutral setting) and $BA_h(0.9,0.1) \rightarrow BA_h(0.9,min_2)$ (homophilic setting).  We then vary the quota between $min_2=0.1$ and $min_2=0.9$. This corresponds to increasing the final minority size from $10\%$ to $50\%$ ($min_2=2*min_{total}-min_1$). The simulations reveal opposite effects: In the homophilic setting (purple), the minority representation grows with the final minority size and quota. On the contrary, in the case of a fixed heterophilic mixing preference (orange), the final minority representation decreases with rising minority sizes. As a result, the homophilic setting starts to benefit the minority more than the heterophilic setting for high quotas (ii). In the neutral case (grey), we see no impact of a quota at all. From these separate investigations of the two types of interventions, we can see that they are not simply benefiting the minority in all cases, but their impact seems to be dependent on an interaction effect of homophily and group size in the post-intervention phase.}
    \label{fig:2}
\end{figure*}

To evaluate the impact of network growth interventions, we consider rankings based on node degree. 
It has been shown that especially in homophilic networks minorities are less represented in the top-ranked nodes \cite{karimi_homophily_2018}.
As shown in Figure \ref{fig:2}A, we compute the rankings for the pre- and post-intervention networks in the growth process. 
To determine the visibility of the minority group, we examine the minority fraction in the top 100 nodes.
We compare the minority fraction in the top 100 nodes in the pre-intervention network  $BA_h(h_1, min_1)$ (dotted line) to the post-intervention network (solid line) generated by $BA_h(h_2, min_2)$.
The dashed black line in \ref{fig:2}A indicates what would be a proportional representation according to the total minority fraction in the final post-intervention network: if the dotted line is below the dashed line, the minority is under-represented in the pre-intervention network.
The minority remains under-represented in the post-intervention network for all intervention parameter values for which the solid curve is below the dashed line.
Proportional representation of the minority group is achieved as soon as the solid line crosses the dashed line.

We start by examining the effect of group-size and behavioural interventions separately (Figure \ref{fig:2} B). 
In the pre-intervention phase, we generate networks with $N_1=2500$ nodes. 
We fix an initial minority size of $min_1=0.1$ for all settings and create both a strongly heterophilic network ($h_1=0.1$) and a strongly homophilic network ($h_1=0.9$) to compare the influence of the pre-intervention network structure on the effect of the intervention.
We then systematically vary the homophily parameter (behavioural intervention, left) or the minority size (group size intervention, right) and continue the growth process until the size of the network reaches $N=N_1+N2=5000$ nodes. 

\paragraph*{Behavioural intervention.}
To evaluate the impact of behavioural change on minority representation, we fix the total minority fraction of $min_1=min_2=0.1$ and vary $h_2$ between heterophilic ($h_2=0.1$), neutral ($h_2=0.5$) and homophilic ($h_2=0.9$) preferences (Figure \ref{fig:2}B, left).
Starting from a heterophilic pre-intervention network ($h_1=0.1$; orange), the minority is generally over-represented in top-ranked nodes. 
This is enlarged with heterophilic post-intervention values as highlighted in box (i), and the effect decreases as a function of $h_2$. 
Changing to homophilic post-intervention behaviour decreases the minority's visibility in comparison to the pre-intervention baseline. 
This effect is also observable in the homophilic pre-intervention setting ($h_1=0.9$; purple). 
However, in this case the minority is under-represented in the pre-intervention rankings. 
Changing from homophilic to heterophilic behaviour greatly benefits the minority, reversing them from under- to over-representation [see (i)].

\paragraph*{Group size intervention.}
We also examine the individual impact of group size changes, while keeping the mixing preferences $h_1 = h_2$ of the two phases fixed (Figure \ref{fig:2}B, right). 
To this end, we raise the quota, i.e., the minority fraction in arriving nodes in the post-intervention phase, from $min_2=0.1$ to $min_2=0.9$. 
Starting from a pre-intervention setting with $min_1=0.1$ and equal time windows, this corresponds to an increase in the final minority size from $10\%$ to $50\%$ ($min_2=2*min_{total}-min_1$). 
The simulations reveal contrary effects depending on the underlying attachment behaviour of the nodes: In the homophilic setting, the minority representation grows with the final minority size and quota. 
In contrast, in the case of a fixed heterophilic mixing preference, the final minority representation actually decreases when raising the minority size. 
As a result, the minority benefits more from homophilic behaviour for high quotas (ii).

Our results emphasise that the minority does not benefit from all interventions, and there appears to be an interaction effect of homophily and group size in the post-intervention phase. Moreover, the baseline of change is dependent on the pre-intervention homophily in the system, which is significantly lower for homophilic settings, meaning that the representation of a minority in the rankings of an initially homophilic setting is harder to change.
To better evaluate this interdependence of the group size and behavioural interventions, we examine their joint effect in the following.

\begin{figure*}
    \centering
    \begin{subfigure}[b]{\textwidth}
        \centering
        \includegraphics[width=\textwidth]{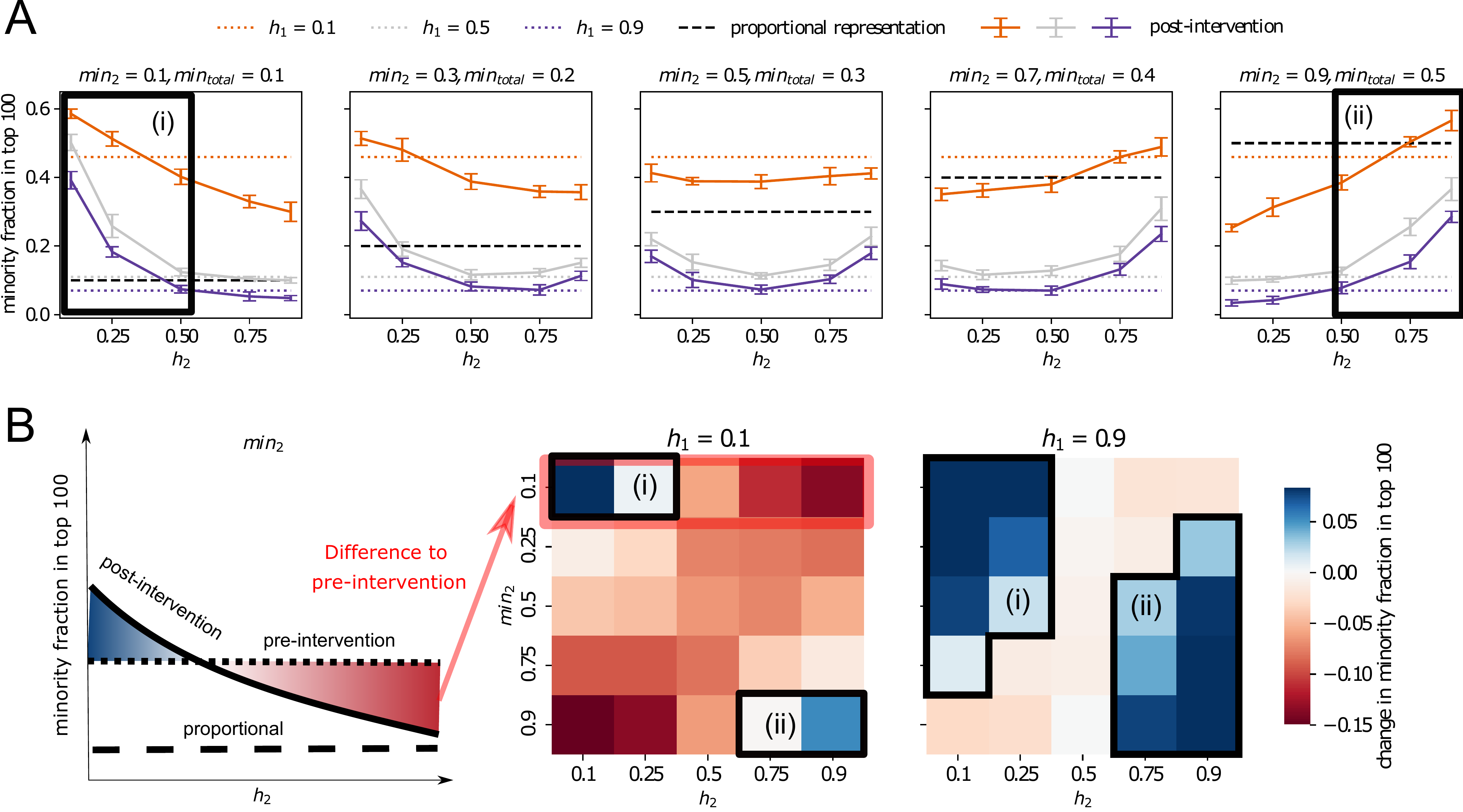}
    \end{subfigure}
        \caption{\textbf{Interaction effect of behavioural and group size interventions.} \textbf{A} We examine $BA_h(0.1,0.1) \rightarrow BA_h(h_2,min_2)$, $BA_h(0.5,0.1) \rightarrow BA_h(h_2,min_2)$ and $BA_h(0.9,0.1) \rightarrow BA_h(h_2,min_2)$, i.e. network growth interventions with an initial minority size of $min_1=0.1$ and heterophilic, neutral  and homophilic pre-intervention behaviour (horizontal dotted lines). We vary both the post-intervention behaviour $h_2$ and the quota $min_2$ to examine the interaction of the two dimensions of interventions. When evaluating which behavioural intervention advances the minority most in terms of ranking visibility, we observe a qualitative shift with increasing quota (from left to right panel). For $min_2 < 0.5$, heterophilic behaviour is more effective (i), for $min_2 >0.5$ homophilic behaviour should be adopted by the minority (ii). This shift occurs in all pre-intervention settings. In the case of $min_2=0.5$ (centre), both heterophilic and homophilic attachment serve the minority equally, but both are better than random attachment ($h_2=0.5$).  \textbf{B} shows the difference of the minority fraction in top ranks in the pre and post-intervention. Red shaded colours indicate that the minority has decreased their representation compared to the pre-intervention network, blue shades indicate an improvement in visibility. For both heterophilic (left panel) and homophilic (right panel) pre-intervention networks, when evaluating which intervention advances the visibility of the minority most, there is a shift from heterophilic (i) to homophilic attachment (ii) for growing quotas, equivalently to A. Additionally, one can see that for homophilic initial settings (right) the improvement of the minority visibility is stronger, whereas for heterophilic initial settings the minority loses their initial (over-)representation for wider parameter ranges. These results show that behavioural and group size changes must be coordinated to achieve a beneficial effect on the visibility of the minority. }
        \label{fig:3}
\end{figure*}

\paragraph{Combined intervention.} 
To investigate the possible interaction of group-size and behavioural interventions, we vary both the post-intervention behaviour $h_2$ and the quota $min_2$ in Figure \ref{fig:3}. 
When evaluating which behavioural intervention advances the minority most in ranking visibility, we observe a qualitative shift with increasing quota in Figure \ref{fig:3}A.
For $min_2 < 0.5$, heterophilic behaviour is more beneficial (i), whereas for $min_2 >0.5$ homophilic behaviour (ii) should be adopted to further the minority's visibility.
This shift can be observed for all three (heterophilic, neutral and homophilic) pre-intervention settings. 
In the case of $min_2=0.5$ (centre), both an increase in heterophilic and homophilic post-intervention behaviour serve the minority equally, and both are better than random attachment ($h_2=0.5$). 
In this case, random behaviour without an attachment preference becomes the worst type of behaviour to advance the minority's visibility. 
The results emphasise that interventions cannot be assessed in isolation. 
Behavioural and group-size interventions must be coordinated to improve the visibility of a minority.

In Figure \ref{fig:3}B, we examine the difference of the minority fraction in the pre- and post-intervention rankings, as an indication for how much the intervention increases or decreases the minority visibility, compared to the pre-intervention status.
Red shaded colours indicate that the minority has decreased their visibility, and blue shades indicate an improvement. 
We can see that for both heterophilic (left panel) and homophilic (right panel) pre-intervention networks, there is a shift from heterophilic (i) to homophilic attachment (ii) for growing quotas, in agreement with Figure \ref{fig:3} A. 
Additionally, for homophilic initial settings the improvement of the minority visibility is stronger (right), whereas for heterophilic initial settings the minority loses their initial (over-)representation for wider parameter ranges. 
This indicates that the strength of the intervention impact is highly dependent on the homophily in the pre-intervention network.

\subsection*{Impact of network interventions on the degree dynamics of minority and majority group}
We hypothesise that the observed shift in Figure \ref{fig:3} occurs due to a role change in the post-intervention phase: If $min_2 > 0.5$, the minority becomes the majority in the set of newly arriving nodes in the post-intervention phase. 
To improve the minority's visibility as a group, homophilic behaviour that originally benefited the (former) majority should be adopted. 
We now evaluate further if this role change is indeed the reason for the observed shift. 

\begin{figure*}
    \centering
    \begin{subfigure}[b]{\textwidth}
        \centering
        \includegraphics[width=\textwidth]{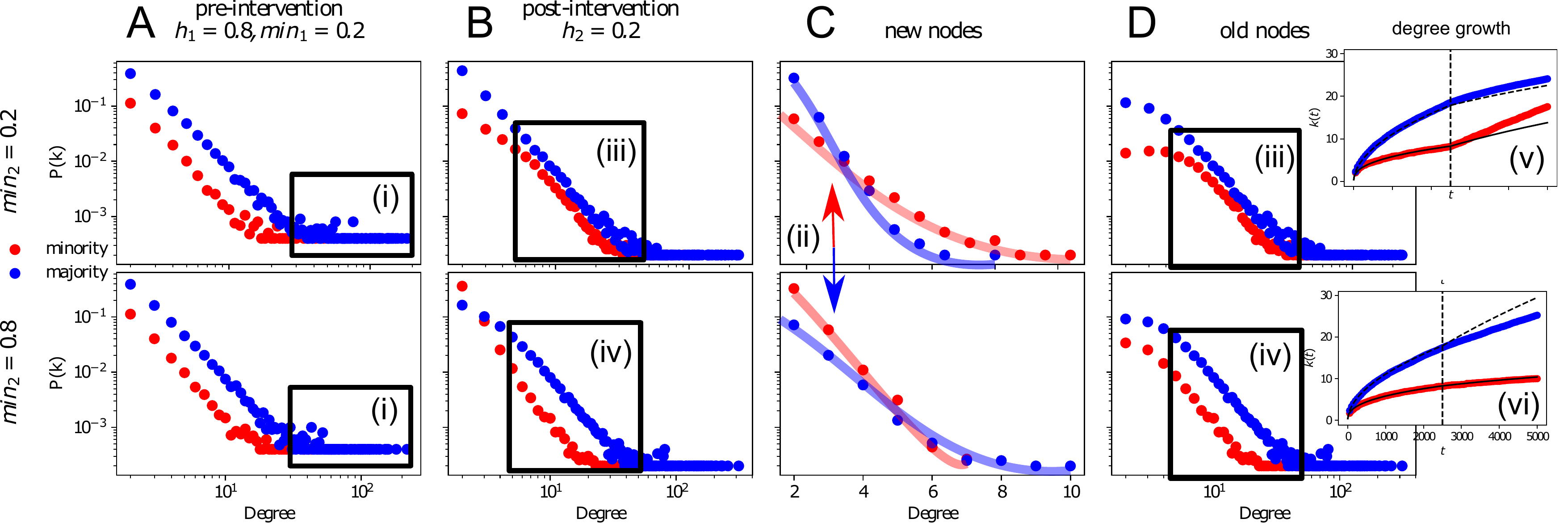}
    \end{subfigure}
        \caption{\textbf{Degree distribution and degree growth of  $BA_h(0.8,0.2) \rightarrow BA_h(0.2,min_2)$.} We examine the degree distributions of the pre and post-intervention network of an intervention, given by $BA_h(0.8,0.2) \rightarrow BA_h(0.2,min_2)$ (switch from homophilic to heterophilic behaviour for different post-intervention minority sizes). In Figure \ref{fig:4} A we can see the degree distributions of the pre-intervention networks, which are the same for settings with and without group intervention (both rows) as we start the interventions from the same pre-intervention network. Due to its homophilic nature, the majority nodes dominate the high-degree nodes in this case (i). This is in agreement with the classical BA-homophily model \cite{karimi_homophily_2018}. In Figure \ref{fig:4} B, we show the degree distribution of the final network after the post-intervention phase. To examine the effects more closely, we split up the nodes in two groups, those that arrived after the intervention (new nodes, Figure \ref{fig:4} C) and those that arrived prior to the intervention (old nodes, Figure \ref{fig:4} D). In Figure \ref{fig:4} C we see that the newly arriving nodes show a degree distribution in which the minority nodes dominate the high-degree nodes if no quota is applied ($min_1=min_2=0.2$, top row). 
However, in the case when an $80\%$ quota is applied (bottom row), the majority group forms the numerical minority of the newly arriving nodes in the post-intervention phase. 
Therefore, the roles of the two groups are interchanged (ii) and the majority, now in the minority in the newly arriving nodes, dominates the high-degree nodes. This explains the interaction effect of homophily and quota parameters in Figure \ref{fig:3}, as the minority should now adopt homophilic attachment as soon as they are the majority of the newly arriving nodes, i.e. for $min_2>0.5$. 
This role change only shows such a strong impact for the degree distributions of the newly arriving nodes. 
In contrast, in Figure \ref{fig:4}  B and D we  see that the intervention only has a weak effect on the degree distribution of the old nodes and thus also the final network. This is because the preferential attachment mechanism in the initial phase leads to ranking stability of the high-degree nodes \cite{ghoshal_ranking_2011} and their position in top ranks can not easily be replaced.
We can only find a small increase of higher degree probability for the minority group (iii) in the case of no quota, which does not lead to overtaking the dominating position of the majority. 
To analyse the size of the effect on the old nodes in more detail, we look at the degree growth of the two groups (insets): The minority shows a small increase in growth for the post-intervention phase (v). This effect is small but it is still impactful for the rankings, as we have seen in in Figure \ref{fig:3}. In contrast, in the case of an $80\%$ quota (bottom), the minority now holds the majority of the newly incoming nodes. Therefore, heterophilic attachment is not beneficial anymore and the intervention has no effect at all (iv). We can also see this in the degree growth plot (vi). More detailed investigations and analytical results for the degree growth can be found in Section \ref{sec:methods} and Figure \ref{fig:5}. Our detailed model investigations show that new nodes that arrive after the intervention develop their degree clearly according to the new growth parameters, whereas the old nodes are affected less due to their degrees' dependency on the first growth phase. Certain parameter combinations can result in no effect. This emphasises the need for careful consideration of the right combinations of behavioural and group size interventions to impact the minority in the network at all.}
\label{fig:4}
\end{figure*}

We analyse the degree distributions of the pre- and post-intervention networks in Figure \ref{fig:4}. 
As an illustration, we consider the example of a behavioural intervention, switching from homophilic to heterophilic behaviour $BA_h(0.8,0.2) \rightarrow BA_h(0.2,min_2)$. 
A complete collection of all cases can be found in the the Supplementary Information (SI). 
We examine this behavioural intervention combined with no group size intervention (fixing $min_1=min_2 = 0.2$, top row) and combined with a quota of $min_2=0.8$ (bottom row). 
This quota results in a network with parity, meaning a final minority size of $50\%$. 

In Figure \ref{fig:4} A we see the degree distributions of the pre-intervention networks, which are the same for settings with and without group intervention (both rows) as we start the interventions from the same pre-intervention network. 
Due to its homophilic nature, the majority nodes dominate the high-degree nodes (i), in agreement with the classical Barab\'{a}si-Albert-homophily model \cite{karimi_homophily_2018}. 

In Figure \ref{fig:4} B, we show the degree distribution of the final network after the post-intervention phase. 
To examine the effects more closely, we can split the nodes up in the group that arrived after the intervention (new nodes, Figure \ref{fig:4} C) and before to the intervention (old nodes, Figure \ref{fig:4} D). 

In Figure \ref{fig:4} C we can see that the newly arriving nodes show a degree distribution in which the minority nodes dominate the high-degree nodes if no quota is applied ($min_1=min_2=0.2$, top row). This agrees with the classical Barab\'{a}si-Albert-homophily model \cite{karimi_homophily_2018} for heterophilic attachment, which is the post-intervention behaviour in this case.
However, in the case when an $80\%$ quota is applied (bottom row), the majority group forms the numerical minority of the newly arriving nodes in the post-intervention phase. 
Therefore, the roles of the two groups are interchanged (ii) and the majority, now in the minority in the newly arriving nodes, dominates the high-degree nodes. This explains the interaction effect of homophily and quota parameters in Figure \ref{fig:3}, as the minority should adopt majority-favouring behaviour (i.e. homophilic attachment) as soon as they are the majority of the newly arriving nodes, i.e. for $min_2>0.5$. 

This role change only shows such a strong impact for the degree distributions of the newly arriving nodes. 
In contrast, in Figure \ref{fig:4}  B and D we can see that the intervention only has a weak effect on the degree distribution of the old nodes and thus also the final network. This is because the preferential attachment mechanism in $BA_h(h_1,min_1)$ leads to ranking stability of the high-degree nodes \cite{ghoshal_ranking_2011} and their position can thus not easily be changed by $BA_h(h_2,min_2)$.

More specifically, we can only find a small increase of higher degree probability for the minority group (iii) in the case of no quota, which does not lead to overtaking the dominating position of the majority. 
To analyse the size of the effect on the old nodes in more detail, we look at the degree growth of the two groups: The minority shows a small increase in growth for the post-intervention phase (v). 
However, the fact that there is a small increase in high-degree minority nodes in the degree distributions is sufficient to affect the minority representation in the degree-based rankings for certain intervention combinations as we have observed in the shift in the ranking results in Figure \ref{fig:3}.
In contrast, in the case of an $80\%$ quota (bottom), the minority now holds the majority of the newly arriving nodes. Therefore, heterophilic attachment is not beneficial for the minority anymore to receive a lot of new links and the intervention has no effect at all (iv).
We can also see this in the degree growth plot (vi). 
More detailed investigations and analytical results for the degree growth can be found in Section \ref{sec:methods} and Figure \ref{fig:5}.
Our detailed model investigations show that new nodes that arrive after the intervention develop their degree clearly according to the new growth parameters, whereas the old nodes are affected less due to their degrees' dependency on the first growth phase.
Certain parameter combinations can result in no effect. 
This emphasises the need for careful consideration of the right combinations of behavioural and group size interventions to impact the minority in the network.

\subsection*{Distance from proportional representation}
\begin{figure}
    \centering
    \includegraphics[width=0.8\textwidth]{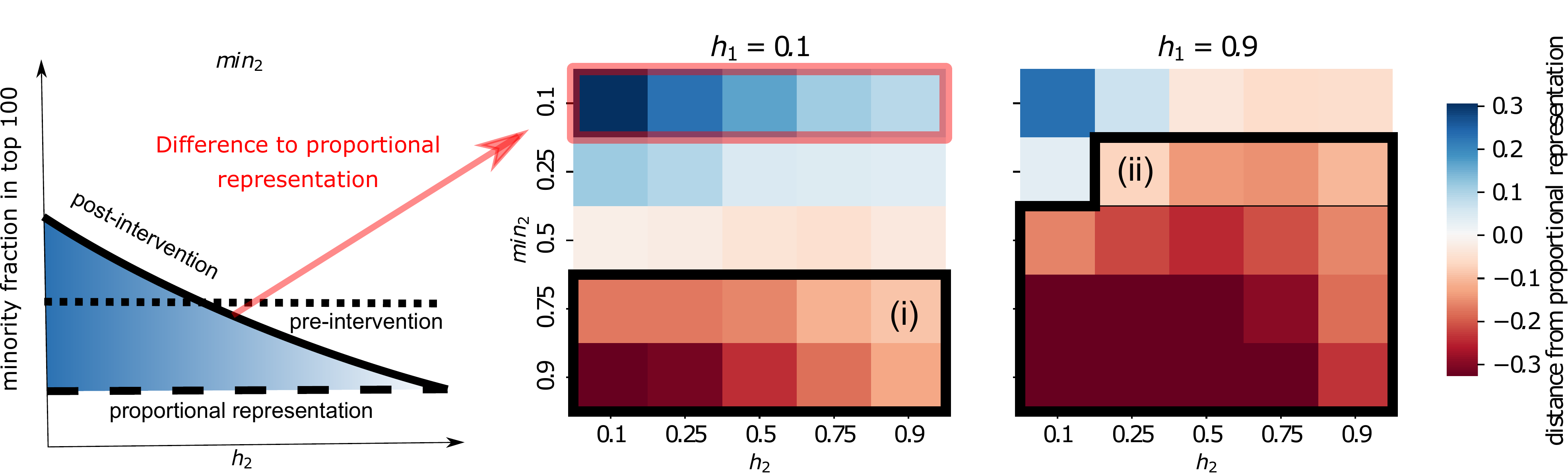}
    \caption{\textbf{Distance of post-intervention ranking visibility to proportional representation.} We examine the difference between the minority fraction in the post-intervention rankings and a proportional representation. A proportional representation would exist if the minority fraction in the top 100 nodes corresponds to the total size of the group in the network. Red-shaded colours indicate that the minority is under-represented concerning their total size, the blue-shaded colours indicate over-representation in the top 100 ranked nodes. We can see that for a homophilic pre-intervention setting (right), this distance to proportional representation is more pronounced for a wide range of intervention parameter combinations (ii) than for an initial heterophilic setting, where proportional representation can only not be achieved for very high quotas resulting in high total minority sizes (i), This underlines the fact that the pre-intervention network highly determines the potential benefit of forward-looking network growth interventions. If the initial setting is determining a strong underrepresentation of the minority, this can hardly be changed by adapting the network growth. Moreover, we observe that the distance to proportional representation increases with increasing quota as the quota raises the final minority size, so also the minority fraction in the proportional representation. However, despite bringing more minority nodes into the system, these new nodes do not appear in the top degree-ranked nodes due to the structural inequality that results from the pre-intervention network. This underlines that in the case of very homophilic pre-intervention networks, interventions that change the future growth of a network are not enough to achieve proportional representation in rankings. }
    \label{fig:6}
\end{figure}

As a separate analysis, we examine the difference of the minority fraction in the post-intervention rankings to a proportional representation in Figure \ref{fig:6}.
A proportional representation would exist if the minority fraction in the top 100 nodes corresponds to the total relative size of the group in the network. 
This fraction is here given by the final minority fraction $min_{final}=(min_2+min_1)/2$. 
Red-shaded colours indicate that the minority is under-represented concerning their total relative size, and blue-shaded colours indicate over-representation in the top 100 ranked nodes.
It is interesting to see that the possibility of reaching a proportional representation is highly dependent on the pre-intervention homophily: In the case of a homophilic pre-intervention network (right), the distance to proportional representation is much higher for a wider range of intervention parameters [see (ii)] than for heterophilic pre-intervention graphs [see (i)]. 
This highlights again that there is a huge dependency of the high-degree nodes on the pre-intervention growth process. 
This is a property of scale-free networks arising from preferential attachment, which exhibit a very stable ranking order due to stable degree sequences for high-degree nodes \cite{ghoshal_ranking_2011}. 
It emphasises that the pre-intervention network highly determines the potential benefit of forward-looking network growth interventions. 
If the initial setting results in a strong structural disadvantage of the minority (homophilic pre-intervention setting), this can hardly be changed by small adaptations in the network growth. 
These observations suggest, that proportional representations might not be possible to achieve only by interventions which change the future growth process of the network, as the interventions are heavily influenced by the structural inequality already encoded in the pre-intervention network. Therefore it might be necessary to alter the structure of the initial network in addition to influencing its future growth.
Moreover, we observe that the distance to proportional representation increases with increasing quota.
The quota brings more minority nodes into the system but due to the structural inequality encoded in the pre-intervention network, these new nodes do not appear in the top degree-ranked nodes on short time scales. 

\subsection*{Network interventions in real collaboration networks}

\begin{figure}
    \centering
    \includegraphics[width=\textwidth]{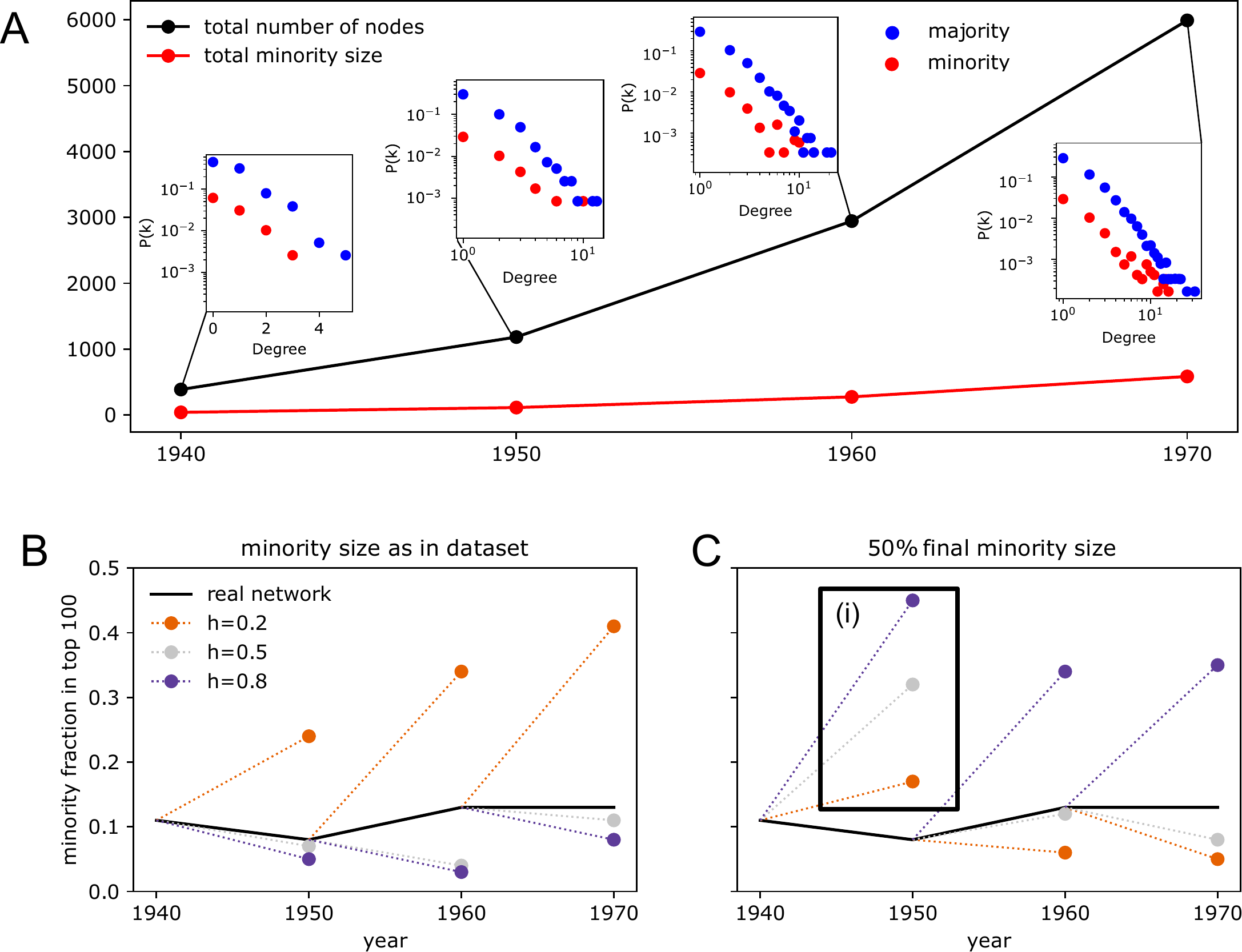}
    \caption{\textbf{Impact of network growth interventions on the collaboration network}. We examine the potential impact of interventions on the representation of women (a minority group in physics) in the top 100 degree nodes in a scientific collaboration network based on publications of the American Physical Society (APS). We aim to evaluate, which combination of interventions would have been effective for this example of a real-world setting. Panel \textbf{A} shows the total number of nodes and the total minority size of the collaboration network in a period of 30 years, including the degree distributions of minority and majority nodes every 10 years. We see no significant increase in the minority size in the real network. In panel B and C we simulate a range of interventions using the original networks as the starting points of post-intervention growth, which is performed until the synthetic network reaches the size of the real network after $10$ years. We then compare the final ranking of minorities based on the intervention with the one of the real network at that point in time. In \textbf{B}, we fix the final minority size as in the real networks and only consider behavioural interventions. In \textbf{C} we apply a minority quota such that the final minority size is raised to $50\%$ each $10$ years. In the case without a minority quota in B, we observe that female scientists quickly achieve a better representation for a heterophilic behavioural intervention ($h_2=0.2$). In contrast, in C we observe a significant positive effect of behavioural intervention. With a group size quota, homophilic behaviour improves the visibility of the minorities most ($h_2=0.8$). This agrees with the results of our model in Figure \ref{fig:3}. However, it is interesting to see that for the first network in 1940, all types of behaviour lead to an improvement simply by installing a quota (i). This is because the 1940 network has not established the scale-free property sufficiently yet, which one can observe in the degree distribution of the 1940 network in A. The high degree nodes are not as stable yet and the newly arriving group of minority nodes thus still have a chance of becoming part of that group from early on. This indicates that the point of time of installing a quota has a significant impact on its potential benefit, especially if the preferential attachment mechanism has not established yet. This effect is irrelevant for pure behaviour interventions (B). Our results show that only specific combinations of behavioural and group size interventions can increase the minority's visibility in real-world networks and specifically for quotas, early timing is essential.}
    \label{fig:aps_minority_fraction}
\end{figure}

Exploring the impact of network growth interventions in real-world systems can inform how to improve the visibility of a minority therein.
Here, we consider the example of scientific collaborations in works published by the American Physical Society (APS), a setting in which female scientists are historically under-represented \cite{kong_first-mover_2021}. We construct a temporally growing co-authorship network of one generation by linking authors that collaborated on a paper published in one of the APS journals between $1940$ and $1970$. We consider the attribute gender, whereas women form the numerical minority. We acknowledge that the notion of gender is fluid and non-binary, and here we only consider gender as a binary attribute in terms of social perception of names~\cite{kong_first-mover_2021}. A full description of the data set and the used methodologies can be found in the SI. 

In Figure \ref{fig:aps_minority_fraction}A, we plot the total number of nodes and the total minority size of the collaboration network in a period of 30 years, including the degree distributions of minority and majority nodes every 10 years.
We see that there is no strong increase in the minority size in the real network, which makes this an interesting case for hypothetical intervention scenarios which we can simulate with our model. 

More specifically, group size interventions (i.e. more women joining the academic job market) and behavioural interventions (i.e. more or less diverse collaborations) can occur in the future. For example, policymakers might resort to affirmative action and enforce a quota in the hiring processes of universities or the minority group themselves might grow naturally due to a rising number of female university graduates in Physics. 
Also, policymakers or administrators could further support behavioural change by tuning the mixing of scientific groups, e.g., when composing teams or at scientific events, such as conferences or schools.
Our model allows us to study the representational impact of both behavioural and group-size interventions and their interplay in a systematic way.
 
To this end, we simulate a range of hypothetical interventions using the original networks as the starting points of a post-intervention growth process $BA_h(h_2,min_2)$, which is performed until the synthetic network reaches the size of the real network after $10$ years. 
We then compare the final minority representation in the network where we applied an intervention with the one of the real network at that time point. 

In Figure \ref{fig:aps_minority_fraction} B, we first consider pure behavioural interventions by fixing the final relative minority size to the one in the real network. 
In alignment with the results of the synthetic simulations in Figure \ref{fig:3}, we find that the minority can improve their representation only if the behavioural intervention is heterophilic. 
In this case, the minority receives more links due to their group size, as all of nodes of the larger majority group have a preference of attaching to the minority nodes.
In other words, the groups must collaborate, in a heterophilic way ($h_2=0.2$), with each other to improve women's visibility in the top-ranked nodes.
In the case of homophilic ($h_2=0.8$) or random behavioural interventions ($h_2=0.5$), their ranking representation is decreased.

However, when enforcing a final total minority size of $50\%$ after each $10$ years, i.e. equal representation in the system, Figure~\ref{fig:aps_minority_fraction} C shows the expected switch of beneficial behavioural intervention that we already observed in Figure \ref{fig:3}: as soon as women are in the majority of newly arriving nodes, homophilic behavioural interventions become more beneficial ($h_2=0.8$). 
In contrast to the case of no quota in Figure \ref{fig:aps_minority_fraction} B, women now benefit from inter-gender collaboration as they function as the majority among the newly arriving scientists.

Additionally, we observe an interesting timing effect of the quota in Figure \ref{fig:aps_minority_fraction} C: in the early stages of the pre-intervention growth process, a quota is highly beneficial in all cases of behavioural interventions (i). 
This is because the 1940 network is not sufficiently scale-free yet, which means that the preferential attachment mechanism has not yet established super-stable high-degree nodes. 
Therefore, increasing the minority size early on can still very efficiently impact the ranking representation of the minority, as the newly arriving minority nodes can still compete with the high-degree majority nodes that arrived in the pre-intervention phase. 
This underlines the importance of establishing group-size interventions as early as possible in the growth process of the network.

%% file: 3_Discussion.tex
\section{Discussion}
\label{sec:Discussion}
Our findings demonstrate that the outcome of group-size and behavioural interventions cannot be evaluated separately, but are inter-dependent. 
More specifically, our results show that it is not enough to increase the size of a minority group to increase the minority's visibility, but there has to be an additional consideration of behavioural aspects which influence the network structure. 
Without a behavioural adaptation, it can become virtually impossible to reach proportional representations in rankings, despite extreme quotas. This underlines the necessity of behavioural interventions such as increasing networking in minority groups if they are large enough to accumulate a cumulative advantage in a growing social network. 
On the other hand, if quotas are not sufficiently large, heterophilic mixing should be encouraged, because the minority will not increase their visibility without connecting to the majority group. Our analytical derivations enable us to explicitly find the intervention combinations which have sufficient impact on the growth of the high-degree nodes, and thus on the minority representation in the rankings.
Additionally, there is a high dependency of the interventions on the initial conditions of the network: If the initial configuration is strongly homophilic, even a very strong group size intervention cannot improve the minorities' representation to a level which is proportional to their total size due to the ranking stability of the high degree nodes \cite{ghoshal_ranking_2011}. 
Minorities are then locked into their initial network position due to historical and cumulative structural inequality. 


It is important to emphasise that we use the notion of minority to refer to a group of people who share a similar attribute (e.g., gender, race, ethnicity) and whose numeric size is smaller than other groups. This imbalance in population size often, but not always, results in inequalities and marginalization. Marginalization is defined as ``to relegate to an unimportant or powerless position within a society or group". The terms marginalised and minority groups are often used as equivalents, but as we have shown earlier, a majority group can be marginalised (heterophilic setting) or advantaged (homophilic setting) in terms of their network position. 
Therefore, marginalisation of a group does not necessarily correlate with the group size. 
In our setting, marginalisation measured by ranking position is the implication of a joint effect of homophily and group size. 
We are aware that this is only one aspect of measuring marginalisation and future work could address other ways to examine inequality in terms of other network measures, impact on algorithmic decision making, and policy decisions to only name a few. 

Moreover, here we are discussing groups concerning certain binary labels such as gender or race. 
However, this does not imply that the attributes we are considering are themselves binary.
As an illustration, consider minority and majority groups concerning gender. 
We acknowledge that the notion of gender is fluid and non-binary, but in many social contexts such as STEM academia, the group of people identifying with non-binary, trans and female gender is smaller than the group of cis men and would therefore form a minority group in our model. 
In future work one can extend the binary attribute setting to continuous and categorical variables and also look at higher-dimensional attributes to consider the effect of interventions concerning different attribute combinations.

Additionally, it is important to note that in our work we sometimes simulate extreme parameter ranges (i.e. minorities connect only with minorities which may be unrealistic in a practical setting). 
We do not suggest these extremes as the policy intervention to take. 
However, they can show that even unrealistically extreme scenarios, e.g., very high quotas would not be enough to fulfil improvements to the intended outcome if they do not align with behaviour. 
The main point of these simulations is thus to show the qualitative interaction effect of behavioural and group size interventions. This should encourage policymakers to think more than one-dimensional when considering measures that combat structural inequality. 
It is not meant to derive exact parameter choices for policies, which would not be practical in an application in any case as even very specific policies are often influenced by the adaptation of the individuals (when trying to enhance certain behaviour) or institutions (in the case of quotas). 
Moreover, we measure the impact of interventions on the visibility of the minority here, but inequality has more facets than just that. For example, support networks despite of not having a big effect on institutional inequality, could provide significant support in the individual career development and community building of their members \cite{dennissen_diversity_2019}. 
These aspects of behavioural interventions are very necessary to consider and it is important to emphasise, that our work provides a rigorous analysis of why these interventions can sometimes not improve other aspects of inequality and how they thus should be extended, not replaced, by other measures. The proposed two-phase network model, despite being able to capture realistic ranking of groups \cite{karimi_homophily_2018}, do not capture all types of micro-level social mechanisms such as triadic closure or higher-order interactions. However, we believe that distilling the concept of homophily in such a simple model enables us to understand the isolated mechanism of homophily and group size on representation more clearly and are still representative of typical social networks. 

Overall, our analysis enables us to evaluate the effectiveness of possible interventions that aim to change structural inequalities in social networks. 
We find that even extreme group size interventions do not have a strong effect on the position of minorities in rankings if certain behavioural interventions do not manifest at the same time. For example, minority representation in rankings is not increased by quotas if the network does not additionally adopt an appropriate behaviour. 
Moreover, the effectiveness of the interventions is highly dependent on the initial conditions in social networks. 
This emphasises that network-based inequalities are hard to combat only by enforcing quota-based methods without structural considerations. Our work lays a theoretical foundation for a new generation of studies aiming to explore the effectiveness of network interventions computationally and have applications in informing policies in society and algorithms. 

%% file: 4_Methods.tex
\section{Methods}
\label{sec:methods}

\subsection{Network model with time dependant minority size and homophily}
We build our model based on the \textit{BA-Homophily}~\cite{karimi_homophily_2018} attributed network model which is an extension of the preferential attachment model (Barab\'{a}si-Albert \cite{barabasi_emergence_1999}) including a tunable homophily parameter that regulates the mixing probability between two groups $a$ and $b$. 
We consider two groups that differ in their size, a minority and a majority group, and define the fraction of the minority as $f_a= min$. 
We thus have that the relative majority group size is given by $f_b=1-min$. 
We define a homophily parameter $h$ which modulates the probability that two nodes of the same group attach to each other and not to the respective other group.
In general, the homophily values might be different for the majority and minority groups, but we are considering symmetric homophily here for sake of simplicity ($h_{aa} =h_{bb} =h$). 
The inter-group mixing probabilities is the complementary probability: $h_{ab}=h_{ba}=1-h$.
With these two parameters, the growth process is defined by $BA_h(h,min)$ where the probability of an arrival node $j$ to connect to an existing node $i$, $\Pi_{i}$, is given by:
$$
\begin{aligned}
\Pi_{i} = \frac{h_{ij} k_{i}}{\sum_{l} h_{lj} k_{l}}
\end{aligned}
$$
where $k_{i}$ is the degree of node $i$ and $h_{ij}$ is the homophily between nodes $i$ and $j$ \cite{karimi_homophily_2018}. 

We extend this model by considering time-dependent group sizes $min(t)$ and homophily parameters $h(t)$. 
In a growth period  $[0,N]$, in which a total number of $N$ nodes join the network uniformly, one in each time step, the network growth is then given by $BA_h(min(t),h(t))$. 
The parameters can change either continuously or discretely. 
In this work, we consider one change at a discrete time event, the intervention time point $T$. 
We thus have for $t\in[0,N]$
$$
\begin{aligned}
BA_h(min(t),h(t))= \begin{cases}
BA_h(min_1,h_1)& t < T \\
BA_h(min_2,h_2) & \, t\geq T
\end{cases}
\end{aligned}
$$

This creates a two-phase process with a pre- and post-intervention phase.
The time point $T$ clearly determines, that $N_1=T$ nodes join the network in the pre-intervention phase, given by $BA_h(min_1,h_1)$. 
Starting with the network at the time of intervention resulting from this process, the other $N_2=N-T$ nodes in the second phase attach according to $BA_h(min_2,h_2)$. 
A network growth intervention is thus fully defined by $BA_h(h_1,min_1) \rightarrow BA_h(h_2,min_2)$. 
In this work, we consider the two phases to have the same length for simplicity. 
We therefore have that $T=N_1=N_2 =N/2$.  
In the new, two-phase model the probability of an arrival node $j$ to connect to an existing node $i$ now depends additionally on the time point of arrival of the nodes.
$$
\begin{aligned}
\Pi_{i}(t) =  \begin{cases}
\frac{(h_1)_{ij} k_{i}}{\sum_{l} (h_1)_{lj} k_{l}}& t < T \\
\frac{(h_2)_{ij} k_{i}}{\sum_{l} (h_2)_{lj} k_{l}} & \, t\geq T
\end{cases}
\label{eq:homophilic_BA_time}
\end{aligned}
$$

However, note that the node degrees $k_i$ of a node $i$ which already arrived in the pre-intervention phase have first developed according to the first-phase growth law. 
As the attachment probability for $t\geq T$ is dependent on $k_i$, we thus have a dependency of the second on the first phase. 
This will show in the degree growth of the two phases, which we will analyse in the next section.
 
\subsection*{Degree dynamics of the groups in the two-phase model}
The degree growth of the two-phase network model can be distinguished into the degree-growth in the pre- and in the post-intervention.
The degree growth of the minority (or majority) in the first phase is given by the growth process described in the original model \cite{karimi_homophily_2018} and is a function of group size $min_1$, homophily $h_1$, and preferential attachment. 
However, the post-intervention phase results in a degree-growth which additionally depends on the pre-intervention phase, as the high-degree nodes developed in that phase impact the second phase growth due to the preferential attachment mechanism.

\subsubsection*{Pre-intervention phase degree growth}
As the growth in the pre-intervention phase is determined purely by $BA_h(min_1,h_1)$, we can follow the analytical derivations in \cite{karimi_homophily_2018} to obtain an analytical solution for the degree growth. 
Let $K_a(t)$ and $K_b(t)$ be the sum of the degrees of nodes for $t <T$ from group $a$ and $b$ respectively. 
Since the overall growth of the network follows a preferential attachment process, we have:
$$
\begin{aligned}
\label{eq:kakb}
    K_a(t) + K_b(t) = K(t) = 2mt
\end{aligned}
$$
where $m$ is the number of new links added to the network at each time step $t$.
We have already defined the relative fraction of group size for each group in the first phase as $f^1_a = min_1$ and $f^1_b=1-min_1$. 
The evolution of $K_a$ and $K_b$ is given by:
$$
\begin{aligned}\label{eq:evol_ka}
\left\{
  \begin{array}{l}
    \dfrac{dK_a}{dt} = m\left(min_1\left(1 + \dfrac{h_1K_a(t)}{h_1K_a(t) + (1-h_1)K_b(t)}\right) + (1-min_1)\dfrac{(1-h_1)K_a(t)}{h_1K_b(t) + (1-h_1)K_a(t)}\right)\\
    \\
   \dfrac{dK_b}{dt} = m\left((1-min_1)\left(1 + \dfrac{h_1K_b(t)}{h_1K_b(t) + (1-h_1)K_a(t)}\right) + min_1\dfrac{(1-h_1)K_b(t)}{h_1K_a(t) + (1-h_1)K_b(t)}\right)\\
  \end{array}
\right.
\end{aligned}
$$
Now, we want to examine the degree of a single node, defined by $k_a$ and $k_b$ for the two groups, respectively. Let us define $Y^1_a(t)$ and $Y^1_b(t)$ as the average degree of a node $a$ and $b$ at time $t$ in the pre-intervention phase, so for $t<T$:
$$
\begin{aligned}
Y^1_{a}(t) &=h_1 K_{a}(t)+(1-h_1) K_{b}(t) =m t\left(h_1 C+(1-h_1)(2-C)\right) 
\end{aligned}
$$
and
$$
\begin{aligned}
Y^1_{b}(t) &=(1-h_1) K_{a}(t)+h_1 K_{b}(t) =m t\left((1-h_1) C+h_1(2-C)\right)
\end{aligned}
$$
where $C$ denotes the average degree growth rate of group $a$ in the first phase, i.e., $\dfrac{dK_a}{dt} = Cm$, and is calculated as follows:
$$
\begin{aligned}\label{eq:C}
    C = min_1\left(1 + \dfrac{h_1C}{h_1C + (1-h_1)(2 - C)}\right) + (1-min_1)\dfrac{(1-h_1)C}{h_1(2 - C) + (1-h_1)C}.
\end{aligned}
$$
The equation for $C$ can be numerically solved and has three real solutions, but only one in the interval $[0,2]$ and thus valid in this case, because the average growth rate cannot be negative or bigger than $2$ in each time step.

For the minority group $a$, we have for $t<T$:
$$
\begin{aligned}
  \dfrac{dk_a}{dt} &= m \left(min_1\dfrac{h_1k_a}{Y^1_a} + (1-min_1)\dfrac{(1-h_1)k_a}{Y^1_b}\right)\\
  &= \dfrac{k_a}{t}\left(\dfrac{min_1 h_1}{h_1C + (1-h_1)(2-C)} + \dfrac{(1-min_1)(1-h_1)}{(1-h_1)C + h_1(2-C)}\right)\\
  &= \dfrac{k_a}{t}\beta_a^1\\
\end{aligned}
$$
$\beta_a^1$ denotes the growth exponent of the pre-intervention phase for group $a$. We can thus write:
$$
\begin{aligned}
  k_a(t) \propto t^{\beta^1_a} \text{ for } t<T.\\
\end{aligned}
$$

Similarly, for group $b$ we derive the growth exponent $\beta^1_b$ as:
$$
\begin{aligned}
  \dfrac{dk_b}{dt} &= m \left((1-min_1)\dfrac{h_1k_b}{Y_b} + min_1\dfrac{(1-h_1)k_b}{Y_a}\right)\\
  &= \dfrac{k_b}{t}\left(\dfrac{ (1-min)h_1}{(1-h_1)C + h_1(2-C)} + \dfrac{min (1-h_1)}{hC +(1-h_1)(2-C)}\right)\\
  &= \dfrac{k_b}{t}\beta^1_b\\
\end{aligned}
$$
and thus:
$$
\begin{aligned}
  k_b(t) \propto t^{\beta^1_b}\\
\end{aligned}
$$
As the growth exponents for the two groups $\beta_a$ and $\beta_b$ depend on the pre-intervention parameters ($\beta_a^1: = \beta_a(min_1,h_1)$ and $\beta_b^1: = \beta_b(min_1,h_1)$), let us generally refer to these degree growth as $k_a^1(t):=k_a(t)(min_1,h_1)$ and $k_b^1(t):=k_b(t)(min_,h_1)$.

We are now able to investigate the degree growth of each group in the pre-intervention phase as a function of homophily and group size at time $t$ which is expressed in $k^1_a(t) \propto t^{\beta^1_a}$ and $k^1_b(t) \propto t^{\beta^1_b}$. 

Note that there is an inverse relation between the exponent of the degree growth and the exponent of the degree distribution ($p(k) \propto k^\gamma$), as follows \cite{karimi_homophily_2018} :
$$
\begin{aligned}
    \gamma = - (\frac{1}{\beta} + 1)
\end{aligned}
$$

Thus, by estimating the degree growth rate, we can derive the exponent of the degree distribution of the groups and vice versa. 

\subsubsection*{Post-intervention phase degree growth}

In the post-intervention phase, we have to distinguish between two groups of nodes, as shown in Figure \ref{fig:4}: the group of "old" nodes, which arrived at time point $t<T$, before the intervention, and the group of "new" node which arrive in the post-intervention phase at $t\geq T$.  The degree growth of the group of new nodes is purely determined by the new, post-intervention growth model  $BA_h(min_2,h_2)$. The degree growth of these nodes is then given by $k_a(min_2,h_2)$ and $k_b(min_2,h_2)$ equivalently to the derivations for the pre-intervention phase, only with the changed homophily value $h_2$ and minority group size $min_2$. In Figure \ref{fig:4}, we can see that the new nodes show a degree distribution purely determined by the new growth $BA_h(min_2,h_2)$.  

In contrast, the degree-growth of the old nodes switches: in the first phase the growth is given by $k_a(min_1,h_1)$ and $k_b(min_1,h_1)$. Afterwards, in the post-intervention phase it is determined by a mixture model combining $BA_h(min_1,h_1)$ and  $BA_h(min_2,h_2)$. This is because the attachment of a node $j$ arriving in the post-intervention phase $(t\geq T)$ to an existing node $i$, given in Equation \ref{eq:homophilic_BA_time}), is dependent on the node degree $k_i$ of node $i$ and the total degree of all other nodes in the system, which in the case of old nodes have been determined by the first growth phase. 

As the early arriving nodes from the first phase will likely be the high-degree nodes that also dominate the top ranks in the degree rankings, it is important to examine the impact of the intervention on their degree growth in more detail.  Let us start with the group of old minority nodes, group $a$. 
If a new node attaches to the system after the intervention time point $T$, it attaches to an old node according to $k_a^1(t)$ with probability $\alpha$ and due to a new node according to probability $1-\alpha$. Following this logic, we can approximate the degree growth of the old nodes as a linear combination of first and second phase degree growth, weighted by $\alpha$ and $1-\alpha$.
Additionally, as the early arriving nodes from the first phase all have high degrees in the second phase due to their early arrival, they will approximately all receive an equal amount of new edges from nodes from the second growth phase, only due to the preferential attachment mechanism. 
Therefore, we can approximate the average degree growth of this subgroup of old nodes for the minority and majority group respectively as  $\dfrac{dK_a}{dt} = m$, so we have that $K_a(t)=K_b(t) = mt$.  
Thus we obtain
$$
Y^2_a(t) = h_2K_a(t)+(1-h_2)K_b(t) = mt = Y^2_b(t)
$$
We now can write the degree growth of the old nodes in the post-intervention phase approximately as 
$$
\begin{aligned}
\label{eq:parameter_a_old}
\frac{d k_{a}}{d t} &=m \left( (1-\alpha)\left(  min_2\frac{h_2k_{a}}{Y^2_{a}}+ (1-min_2) \frac{(1-h_2) k_{a}}{Y^2_{b}} \right)+ \alpha \beta^1_{a} \right) \\
&=\frac{k_{a}}{t} \left(  (1-\alpha)\left( min_2 h_2+ (1-min_2) (1-h_2)\right) + \alpha \beta^1_{a} \right) \\
&= \frac{k_{a}}{t} \beta_a^{old}
\end{aligned}
$$
and thus the old group of minority nodes switches their degree growth according to
$$
k_{a}(t) \propto t^{\beta_{a}^{old}} \text{ for } t \geq T.
$$Similarly, for group b we now have
$$
\begin{aligned}
\label{eq:parameter_b_old}
  \dfrac{dk_b}{dt} &= m \left( (1-\alpha) \left((1-min_2)\dfrac{h_2 k_b}{Y^2_b} + min_2\dfrac{(1-h_2)k_b}{Y^2_a}\right) + \alpha \beta_b^1\right) \\
  &= \dfrac{k_b}{t} \left( (1-\alpha)\left( (1-min_2)h_2+min_2(1-h_2)\right) + \alpha \beta_b^1 \right)\\
  &= \dfrac{k_b}{t}\beta^{old}_b\\
\end{aligned}
$$
and thus the old group of majority nodes switches their degree growth according to
$$
k_{a}(t) \propto t^{\beta_{a}^{old}} \text{ for } t \geq T.
$$
Now let us determine an approximation for the weighting parameter $\alpha$: If a node joins right at the beginning of the post-intervention phase, so at time point $T$, it has a probability $p=1.0$ to attach to a pre-intervention phase node. If it attaches at the end of the post-intervention growth phase, it has a probability of $p=(N-T)/N$, ($p=0.5$ in our case, as the two growth phases have the same length). Therefore, on average a node of the second phase connects to a pre-intervention node with probability $p=(1.0+0.5)/2=0.75$ and to a post-intervention node with probability $1-p=0.25$. Therefore we have that $\alpha = 0.75$ in our simulations. 

By closer examination of the post-intervention growth parameters$ \beta_a^{old} $ and $\beta_b^{old}$ we see that the degree growth of the old nodes in the post-intervention phase is simply given by the fraction of new edges that the two groups receive respectively. 
Accordingly, it becomes irrelevant to consider the effect of preferential attachment in the second phase, as the old nodes from both minority and majority groups have such a high degree that this effect is negligible in comparison. 
This results in notable symmetries in the degree growth process: If we compare two network growth interventions $BA_h(min_1,h_1) \rightarrow BA_h(min_2^1,h_2^1)$ and $BA_h(min_1,h_1) \rightarrow BA_h(min_2^2,h_2^2)$, the post-intervention degree growth of the old nodes is the same if it holds that both
 $$  
 min^1_2 h^1_2+ (1-min^1_2) (1-h^1_2) = min^2_2 h^2_2+ (1-min^2_2) (1-h^2_2) 
 $$
 and 
 $$
(1-min^1_2)h^1_2+min^1_2(1-h^1_2) = (1-min^2_2)h^2_2+min^2_2(1-h^2_2).
 $$


\begin{figure*}
    \centering
        \begin{subfigure}[b]{\textwidth}
        \centering
        \includegraphics[width=\textwidth]{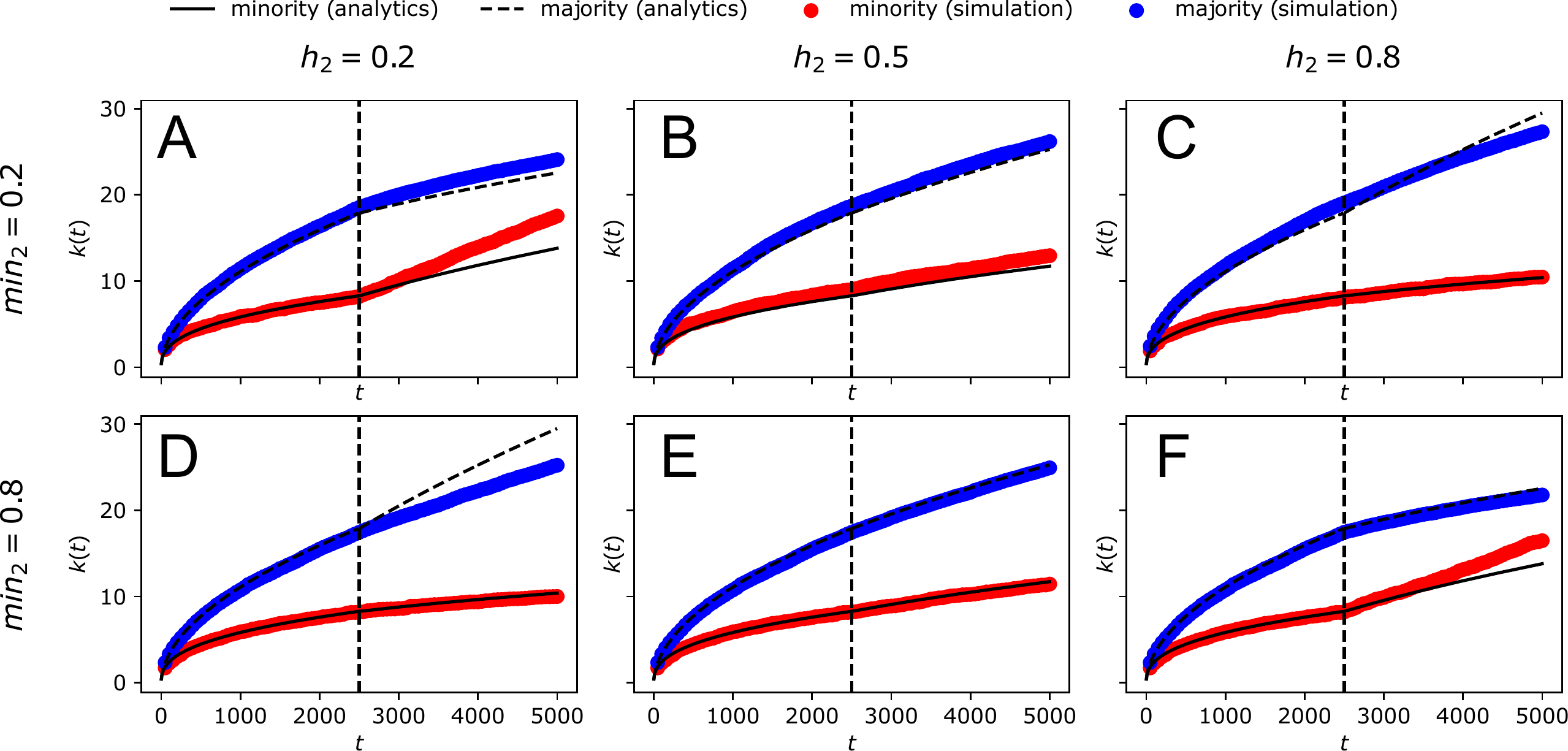}
    \end{subfigure}
        \caption{\textbf{Degree growth of $BA_h(0.8,0.2) \rightarrow BA_h(h_2,min_2)$.} We examine the degree growth of an average minority and majority node from the set of very early arriving nodes from the homophilic pre-intervention phase. We compare our analytical results with the degree growth in the simulations, which generally show a good agreement. Our results show that the interventions generally only show a small effect on the degree growth of the old nodes in the second phase. However, this effect is impactful for the rankings as we have seen in Figure \ref{fig:3}. Moreover, we find that the interaction effect of behavioural and group size intervention parameters creates interesting symmetries in the degree growth: in panel A, we see that in the case of a behavioural intervention from homophilic to heterophilic behaviour but with no group size intervention ($BA_h(0.8,0.2) \rightarrow BA_h(0.2,0.2) $, the degree growth of minority nodes in the post-intervention phase is increased. In panel F, the same effect occurs in a situation with no behavioural intervention but a strong group size intervention ($BA_h(0.8,0.2) \rightarrow BA_h(0.8,0.8) $. Similar symmetries can be observed for panels B and E and for panels C and D. This leads to the interesting situation that applying both behavioural and group size interventions at the same time, as done in panel D, can lead to the effects on the degree growth of the high-degree nodes as not intervening at all as in panel C. These results emphasise the importance of considering the interaction effects of different interventions, as certain combinations can result in similar or no impact on the immediate degree growth of the nodes with a high degree. Moreover, our analytical derivations enable us to explicitly find the intervention combinations which have sufficient impact on the growth of the high-degree nodes, and thus on the minority representation in the rankings. }
        \label{fig:5}
\end{figure*}

We observe these symmetries in Figure  \ref{fig:5} (for a homophilic pre-intervention setting) and the SI (for the other cases). 
We compare our analytical derivations with the degree growth in the simulations and we generally have a good agreement of the analytical and numerical results.
In Figure \ref{fig:5}A, we see that in the case of a behavioural intervention from homophilic to heterophilic behaviour but with no group size intervention ($BA_h(0.8,0.2) \rightarrow BA_h(0.2,0.2) $, the degree growth of minority nodes in the post-intervention phase is increased. 
In panel \ref{fig:5}F, the same effect occurs in a situation with no behavioural intervention but a strong group size intervention ($BA_h(0.8,0.2) \rightarrow BA_h(0.8,0.8) $. 
Even if the effects are small,  they are sufficient to affect the minority representation in the degree-based rankings for these specific intervention combinations, as we have observed in Figure \ref{fig:3}.  
Similar symmetries can be observed for panels \ref{fig:5}B and \ref{fig:5}E and for panels \ref{fig:5}C and \ref{fig:5}D. 
The symmetric intervention impact on the degree growth leads to the interesting situation that applying both behavioural and group size interventions at the same time, as done in panel D, can lead to the effects on the degree growth of the high-degree nodes as not intervening at all as in panel \ref{fig:5}C.  We observe the same symmetry pattern for all pre-intervention settings (see SI). These results emphasise the importance of considering the interaction effects of different interventions, as certain combinations can result in similar or no impact on the immediate degree growth of the nodes with a high degree. They give an explanation for the observation that only certain combinations of behavioural and group size interventions show a positive effect on the representation of minorities in the top ranks of a network in Figure \ref{fig:3}. Our analytical derivations enable us to explicitly find the intervention combinations which have sufficient impact on the growth of the high-degree nodes, and thus on the minority representation in the rankings.